# Unbounded Operators, Physicality, and Modality in Quantum Theories

*Zhonghao Lu*
*Faculty of Physics, Ludwig-Maximilians University of München*
*zhonghao.lu@campus.lmu.de*
*lu_zhonghao@pitt.edu*
*lu_zhonghao@pku.edu.cn*
*https://orcid.org/0000-0002-7564-2652*

**Abstract**

In this paper, I address the problem regarding the physicality of Hilbert space in quantum mechanics, a topic brought into focus by Carcassi *et al*.'s paper "The unphysicality of Hilbert spaces" recently. I argue for the physicality of the Hilbert space in quantum mechanics, and demonstrate that there are inevitable ambiguity and arbitrariness if we want to identify the set of physical states with a proper subset of it. Moreover, I distinguish the ontological and modal reading of physicality from the pragmatic and operational reading of physicality, and then develop what I call the hierarchical view of physicality. According to this view, physicality is not a monolithic concept, but a layered one that admits degrees. I show the connection between the concept of physicality and modal metaphysics, and finally explain how it can be used to address the interpretational problem of quantum field theory brought by the existence of inequivalent representations.

**Keywords**
Quantum mechanics, Hilbert space, Schwartz space, Physicality, Modality in physics

## 1. Introduction

In the textbook formalism of quantum mechanics, we have the following axioms[1]:

[1] I do not imply that these axioms are exhaustive. I have omitted axioms about the symmetry of many-particles system for example, which are not relevant to my

(I) Any possible quantum state of a system is represented by a normed ray $\psi$ of a Hilbert space $\mathcal{H}$ of the system, and *vice versa*.[2]
(II) Any observable of the system is represented by a self-adjoint operator[3] $\boldsymbol{A}$ of the Hilbert space $\mathcal{H}$.
(III) For a state $\psi$, the expectation value of the observable $\boldsymbol{A}$ is $(\psi,\boldsymbol{A}\psi)$.
(IV) If the Hamiltonian of the system is represented by a self-adjoint operator $\boldsymbol{H}$, the dynamic evolution of the quantum state $\psi$ satisfies the Schrödinger's equation $\mathrm{i}\mathrm{d}\psi/\mathrm{d}t=\boldsymbol{H}\psi$.

However, there are problems with the textbook formalism brought about by the fact that most physically significant observables[4] in quantum mechanics, represented by self-adjoint operators, including the position operator $\boldsymbol{x}$, the momentum operator $\boldsymbol{p}$, and most Hamiltonian operators $\boldsymbol{H}$, are unbounded[5]. According to the Hellinger-Toeplitz theorem, an unbounded self-adjoint operator $\boldsymbol{A}$ can only be defined on a subset of the Hilbert space, namely its domain $\mathrm{Dom}(\boldsymbol{A})$. In other words, it cannot be

---

discussions in this paper.

[2] Heathcote [15] puts it in this way: "There is a one-to-one correspondence between the possible states of a system and the normed rays of a Hilbert space $\mathcal{H}$. (p.526)" Strictly speaking, only pure states are represented by normed rays of a Hilbert space in quantum mechanics. If mixed states are included, there is a one-to-one correspondence between the possible states and the density operators of a Hilbert space. Mixed states will only be included in Section 7 in the discussion of the ontology of quantum field theory.

[3] In most of the physics literature, the term "Hermitian operator" is used. Readers are easy to be confused by the different terminologies used in the literature. Most of the physics literature does not provide a strict definition of self-adjoint operators, and their definition of "Hermitian operator" coincides with the definition of "symmetric operator" in the mathematics literature. All self-adjoint operators are symmetric operators, but not *vice versa*. The difference is crucial because the spectral theorem can be only applied to self-adjoint operators, not all symmetric operators. In some literature, however, "Hermitian operator" is used as a synonym of "self-adjoint operator". For clarity, I do not use the term "Hermitian operator" in this paper.

[4] By "physically significant", I mean that these observables are important in physics. I do not imply that those insignificant observables are unphysical, or that they should be dismissed or excluded from the formalism of quantum mechanics.

[5] For a comprehensive introduction to the unbounded self-adjoint operators, see Hall [14].

defined on the whole Hilbert space $\mathcal{H}$.

Therefore, in the axioms of quantum mechanics, (III) and (IV) cannot be applied to states outside the domains Dom($\boldsymbol{A}$) and Dom($\boldsymbol{H}$) respectively, which would raise conceptual difficulties. In the foundations of physics literature, this problem is first noticed by Heathcote in 1990 [15], and then by Carcassi, Calderón, and Aidala in 2025 [3].

Heathcote [15] argues that the inapplicability of (III) and (IV) implies that the axiom (I) should be abandoned. According to Heathcote, only a proper subset of the Hilbert space $\mathcal{H}$ may correspond to possible quantum states, but quantum mechanics fails to specify this subset of physical states. Heathcote boldly argues that quantum mechanics "could be said to be incomplete in the sense that we do not have within the theory an algorithm for generating the appropriate restrictions on $\mathcal{H}$ (p.533)". Heathcote is unable to provide such restrictions, and faces a considerable difficulty that "the domain restrictions on the unbounded operators that one is interested in will not coincide-simply because the domains do not coincide (p.529)". Moreover, Heathcote worries that the operators on the resulting restricted space are no longer self-adjoint, which would further undermine the fruitful formalism of quantum mechanics.

Carcassi *et al*. [3] present the problems in a different fashion. They argue that the Hilbert space structure itself is unphysical, as it contains unphysical states that "require properties or operations that cannot have a physical counterpart". They also try to determine the real set of physical states, as a proper subset of the Hilbert space, which is Heathcote's unfinished task. They tentatively suggest that the Schwartz spaces can fulfil this role, as they have many desirable features. Although they are not convinced that "this is a final and general answer", they believe that "the properties of Schwartz spaces[6] may provide guidance for a possible comprehensive solution". Their proposal has received some criticism and debate [3, 17], but I believe that it has not touched the core of their argument so far.

[6] In quantum mechanics, almost all literature only discusses separable Hilbert spaces. All infinite-dimensional separable Hilbert spaces are isomorphic to each other. Finite dimensional Hilbert spaces are not discussed here, as they do not admit unbounded self-adjoint operators. Therefore, I only use the singular form of "Hilbert space". On the other hand, the Schwartz spaces discussed by Carcassi *et al*. are not isomorphic to each other, and thereby I use the plural form of "Schwartz spaces".

Heathcote and Carcassi *et al*. have used slightly different expressions and arguments in their respective papers, but they face the same foundational issues of quantum mechanics. While Heathcote [15] uses the terms "possible states" and the "incompleteness of theories" more often, these notions are also tightly connected with the existence of counterparts of "physical reality" (p.525). On the other hand, Carcassi *et al*. rarely use the expression "possible states", but are interested in whether certain states are "physical". I suggest that their differing terminologies should not be viewed as referring to entirely disparate concepts. In Sections 7 and 8, I will further elaborate on how the concept of modality (possibility) is closely related to physicality. Furthermore, Carcassi *et al*.'s paper can be read as a supplement to Heathcote's. While Heathcote is uncertain what possible restrictions on the Hilbert space $\mathcal{H}$ would be like, Carcassi *et al*.'s paper presents the Schwartz spaces as an illuminating and inspirational option.

In Sections 2 and 3, I will discuss the problems with the axioms (III) and (IV) respectively. I will review both Heathcote and Carcassi *et al*.'s arguments on the unphysicality of the quantum states outside the domains of the position operator and the Hamiltonian operators. I argue that those states should not be treated as unphysical.

In this paper I am following an ontological or modal reading of physicality. Being physical is a characterisation of possibility. Some states are physical if and only if they are possible by the standards and criteria from physics. This resonates with the definition of physical possibility in modal metaphysics [19]:

> $p$ is physically possible *iff* $p$ is consistent with the laws of nature.
> $p$ is physically necessary *iff* $p$ follows from the laws of nature.

In the above definition in the literature, $p$ is a proposition. It is not difficult to apply the definition of physical possibility to states:

A state $\psi$ is physically possible *iff* for any proposition $p$ holds for $\psi$,

$p$ is consistent with the laws of nature.

In following the ontological or modal reading of physicality, I take "physical possibility" as the synonym of physicality. The laws of nature do not directly follow from the mathematical formalism of physical theories.[7] If we are justified in believing that certain pathological features should be prohibited by nature, certain states that have these features should be classified as unphysical. For example, a spacetime that admits closed causal loops at large scales (such as the Gödel spacetime), despite its consistency with the formalism of general relativity, may violate more fundamental laws of nature regarding causality, and should be considered unphysical. Similarly, certain states in the Hilbert space with pathological properties may be regarded as unphysical.

Carcassi et al.'s paper does not directly take this ontological or modal position. Instead, their paper implies a pragmatic or operational reading of physicality. In Section 4, I will discuss different accounts of physicality and revisit my arguments in Sections 2 and 3. I start firstly with more concrete discussions regarding the physicality of particular states, then elaborate on a generic characterisation of physicality, and finally come back to the more concrete examples.

In Section 5, I will examine the Schwartz spaces as the sets of physical states in quantum mechanics proposed by Carcassi *et al.*. I will reveal the problems with this approach, and argue that these problems demonstrate the inevitable ambiguity and arbitrariness in determining a single set of physically possible states together with a single set of physically possible observables different from the standard axioms (I) and (II). In Section 6, I will discuss some relevant difficulties raised by Heathcote and Lemos [17].

In Section 7, I will provide a more sophisticated account of physicality, which I call the hierarchical view. According to this view, a physical theory determines not a single set of possible worlds, but instead a multiplicity of sets of possible worlds that enable us to define comparable physical possibility. In Section 8, I will use the hierarchical view of physicality developed in Section 7 to address the interpretational problem in quantum field theory.

[7] Loosely speaking, the laws of nature are more objective, while the mathematical formalism of physical theories is theory-dependent.

## 2. The (un)physicality of states outside the domain of the position operator

In this section, I will first consider the problems with states outside the domain of the position operator $\boldsymbol{x}$. Heathcote is content with showing that the axiom (III) fails to hold. I will show that this problem is not a serious one, and we only need some minor revisions to the axiom (III).

In von Neumann's [29] standard formalism of quantum mechanics, each observable is represented by a self-adjoint operator, and can be decomposed via the spectral theorem. For example, for the position operator $\boldsymbol{x}$, we have

$$\boldsymbol{X}=\int\lambda_{\boldsymbol{x}}\mathrm{d}\boldsymbol{P}_{\lambda\boldsymbol{x}}$$

where $\lambda_{\boldsymbol{x}}$'s fall within $\boldsymbol{x}$'s spectra and $\mathrm{d}\boldsymbol{P}_{\lambda\boldsymbol{x}}$'s are their corresponding projection-valued measures that satisfy the normalisation condition

$$\int\mathrm{d}\boldsymbol{P}_{\lambda\boldsymbol{x}}=1.$$

For a system in the quantum state $\psi$, the probability that a single measurement result falls within a Borel measurable set $\varepsilon\in\mathbb{R}$ is

$$(\psi, (\int_{\varepsilon}\mathrm{d}\boldsymbol{P}_{\lambda\boldsymbol{x}})\psi),$$

where $\int_{\varepsilon}\mathrm{d}\boldsymbol{P}_{\lambda\boldsymbol{x}}$ is a projection operator, which is bounded. The above formula is well-defined regardless of whether $\psi$ is within Dom($\boldsymbol{x}$) or not, and therefore we can obtain the probability distribution of measurement results equally well for states outside the domain of the position operator.

Heathcote is fully aware of this fact. However, he insists that states outside Dom($\boldsymbol{x}$) should be excluded from the formalism of quantum mechanics and argues:

> This is effectively just restating the fact that a probability distribution exists for our function $\psi(x)$. However, none of this helps in obtaining an expectation value, since that is given by [$(\varphi, \boldsymbol{x}\psi)= \int\lambda_X(\varphi,\mathrm{d}\boldsymbol{P}_{\lambda\boldsymbol{x}}\psi)$] and is only defined for some [$\psi(x)\in\mathrm{Dom}(\boldsymbol{x})$]. This expectation value, or mean value, is undefined precisely because it is infinite.[8] [15,

[8] I have modified the symbols in square brackets for the sake of consistency.

p.532]

Strictly speaking, being outside Dom($\boldsymbol{x}$) and possessing an infinite expectation value for position are two distinct conditions. If we turn to the coordinate representation, and treat each quantum state $\psi$ not as an element of the abstract Hilbert space, but as a function over $\mathbb{R}$[9] (a "wave function"), we can formally write down the wave function $\boldsymbol{x}\psi(x)$:

$$\boldsymbol{x}\psi(x)=x\psi(x)$$

even if the right side of the formula is not square-integrable. The condition $\psi\notin$Dom($\boldsymbol{x}$), is equivalent to the condition that the following formula

$$\int x^2\psi(x)\overline{\psi}(x)\mathrm{d}x$$

fails to have a finite value in the sense of the Lebesgue integral. On the other hand, the condition that the expectation value of the position operator (in a generalised sense) is infinite is equivalent to the requirement that the following integral

$$\int x\psi(x)\overline{\psi}(x)\mathrm{d}x$$

is infinite. It is possible that a quantum state may fall outside the domain of the position operator but still have a (generalised) expectation value of position. However, I will argue that, even if states have infinite expectation value, they should not be regarded as "unphysical" or impossible.

In quantum mechanics, expectation values of any physical quantities are never directly observed. Each individual observation yields a specific result, and the expectation value is only obtained as the average of long-term observations of an ensemble in the same quantum state. If we cannot obtain an expectation value $<\boldsymbol{x}>_\psi$ or if it is infinite, the long-term average of measurements of $\boldsymbol{x}$ would not converge for an ensemble in the same state $\psi$. This in itself is not a pathological feature. Unless the expectation value $<\boldsymbol{x}>$ enters into in some physically fundamental process, states that fail to have finite expectation values should not be classified as unphysical or impossible.

A contrary example is the Hadamard condition. In semiclassical quantum gravity, the expectation values of energy-momentum field operators appear in its fundamental equation: the semiclassical Einstein

[9] For simplicity, I consider only the one-dimensional case, following Heathcote. It is not difficult to generalise the discussions to higher dimensions.

equations [30, p.86]

$$G=8\pi\langle \boldsymbol{T}\rangle,$$

where $G$ is the Einstein tensor field and $\boldsymbol{T}$ is the energy-momentum operator-valued field, analogous to the classical Einstein equations

$$G=8\pi T.$$

To avoid spacetime singularities, the equation's right-hand side must remain finite[10], and therefore one is tempted to require that all possible physical states in quantum field theory must satisfy the so-called Hadamard condition [11].

I doubt whether similar results would hold for the position operator. While the energy-momentum operator-valued field $\boldsymbol{T}$ is local or "semi-local", the convergence of $\langle \boldsymbol{x}\rangle_\psi$ relies on $\psi$'s behaviour at spatial infinity. When considering the behaviour of the wave function restricted to a bounded spatial region associated with a projection operator $\boldsymbol{P}$, the wave function's expectation value of the position within the region, as calculated by $\langle \boldsymbol{x}\rangle_{P\psi}$, is always finite. It is unlikely that in a successive more fundamental physical theory, the wave function's behaviour at spatial infinity would have a crucial role in its structure. Consequently, it is unlikely that the global expectation value $\langle \boldsymbol{x}\rangle_\psi$ should be restricted, grounded in concerns in physics.

Carcassi *et al*. [3] have more subtle arguments on the unphysicality of states outside the domain of the position operator. They believe that the divergence of the expectation value $\langle \boldsymbol{x}\rangle_\psi$ "clearly" "does not make physical sense". They offer additional arguments to show that that these states are unphysical:

> When we posit that a quantity, like position, has an infinite range of possible values, we simply mean that the observed value can be arbitrarily large, not that we can literally observe an infinite value. This is the difference between potential infinity and actual infinity.
> The argument can be summed up as follows. On physical grounds, we assume the existence of quantities, like position, that can take arbitrarily large values. This requires that, in principle, for each

[10] Strictly speaking, after regularization.

> possible value, there must exist some state whose expectation matches that value. Completeness forces us to include states with an infinite expectation value, which cannot be implemented physically. Thus, the mathematical definition implicitly assumes objects that cannot be physically realized, so the use of completeness for the space of quantum systems is unphysical. [*ibid*., p.5]

This argument continues in the Appendix of their paper:

> Conceptually, we need to treat the infinities coming from unbounded quantities as potential infinities, not actual infinities: we never actually have infinitely many particles or a system positioned at infinity; we have an infinite range of possible values. [*ibid*., p.18]

Furthermore, they consider a group of unitary transformations of the Hilbert space that can turn states with finite expectation values for certain observables (for example, $\boldsymbol{x}^{n}$) into states with infinite expectation values for these observables. The group of unitary transformations can be understood in a passive manner, namely that it merely reflects a change of coordinates, or in an active manner. Carcassi *et al*. argue that these states are physically inequivalent as only the latter admit actual infinities. If such transformations are understood passively, following Carcassi *et al*., they are problematic as they identify physically inequivalent states as equivalent. If such transformations are understood actively, namely, as reflecting time evolution, they cause "the expectation value oscillate from finite to infinite in finite time" and are thus "physically untenable" [*ibid*., p.7]. They continue:

> The idea is that the evolution will simply stretch and shrink the wave function intermittently, potentially oscillating expectation values from finite to infinite and *vice versa* in finite time. Clearly, this is not a physically meaningful time evolution, yet it will correspond to a strongly continuous one-parameter unitary group and, by theorem, will even admit a Hamiltonian. [ibid., p.16]

In this section, I will further argue that either the admission of actual infinities is insufficient for dismissing the physicality of quantum states, or there are no such actual infinities of position about which Carcassi *et al*. worry.

First, the existence of actual infinity is usually not a problem of physics. The appearances of actual infinity are actually quite normal in physics, both in theories and in reality. In classical mechanics and in special relativity, the spacetime we inhabit is infinite in scale. In reality, we may indeed inhabit an infinite universe. Many cosmological models admit this kind of infinity, including the Friedmann–Lemaître–Robertson–Walker metric. In both classical mechanics and relativity, any spacetime region contains infinitely many spacetime points. This would hold in further quantum gravity theories as well, unless the spacetime structure is discrete, which remains a possibility, but far from consensus. The total number of particles may be infinite in our world.[11] These actual infinities in reality are *epistemologically possible*[12], not merely *modally possible*. It is possible to avoid all of these actual infinities in theories, but a significant trade-off would be incurred by adopting such a strict finitist position. Many theories would need to be reconstructed and it is unclear whether it would succeed.

What is actually a problem for physics is the existence of *singularities*, which can sometimes be caused by actual infinities. For example, in classical Newtonian mechanics, the dynamics of a system with infinite particles are usually pathologically indefinite. However, the appearances of singularities are not limited to the systems with infinite particles. There is a more concerning example that particles may exit the spacetime at spatial infinity within a finite period of time. A less concerning example is the "Norton's dome", where singular solutions of the dynamic equation exist, rendering the evolution of the system indefinite.[13] These singularities break down the usual ordinary evolutions.

[11] Defining and counting particles is a notorious problem in quantum field theory. I suppose that one can find an appropriate definition for this claim.

[12] In the language of possible worlds in modality metaphysics, a proposition $p$ is epistemologically possible *iff* it is possible that $p$ holds for the actual world, but we have not determined whether that it true. A proposition $p$ is modally possible *iff* it holds in a possible world (accessible to the actual world).

[13] All of these examples can be found in Earman's paper [9].

No such singularities are brought about by the actual infinities of the expectation values of positions. The singularities in classical mechanics can be cured in quantum mechanics as the evolution of a system with a self-adjoint Hamiltonian is always deterministic. Besides, the distributions of measurement results can be given with no singularities as usual. The real singularities would occur in semiclassical quantum gravity when the Hadamard condition is violated, as I have discussed earlier in this section. In that case, the actual infinities of certain quantum observables lead to the singularities of spacetime structure. However, this goes beyond the scope of ordinary quantum mechanics, and we do not have reasons to expect that the position operators would play similar roles for reasons given earlier in this section.

One may argue that the actual infinities might be part of physics, in either theory or reality. Nevertheless, this would not happen at the observational level. Carcassi *et al*. have emphasised that we never "literally observe an infinite value". This is consistent with the standard formalism of quantum mechanics, where we never observe any infinite values. It should be emphasised that *the expectation values of position are not observables in quantum mechanics*. Position itself is an observable in quantum mechanics. The values yielded by observation all fall within its spectra, each being finite. For a state outside the domain of the position operator, the observed value can be arbitrarily large, but still finite. This aligns with Carcassi *et al*.'s definition of potential infinity, not actual infinity.

Therefore, the possibility of "oscillating expectation values from finite to infinite and vice versa in finite time" is unproblematic, whether understood in the passive or active fashion. This bears no resemblance to the phenomenon in classical mechanics that particles may exit the spacetime at spatial infinity and return within a finite period of time. The states outside the domain of the position operator should still be considered physical.

Carcassi *et al*.'s paper has employed a more pragmatic argument against the physicality of states outside the domain of the position operator, which is based on their implicit pragmatic definition of physicality. I will return to this argument in Section 4, after a more comprehensive discussion on the contrast between the ontological account and the pragmatic account of physicality.

## 3. The (un)physicality of states outside the domains of the Hamiltonian operators

The inapplicability of the axiom (IV) of standard quantum mechanics formalism and the unphysicality of states outside the domains of the Hamiltonian operators is only discussed by Heathcote, who states that

> The Hamiltonian operator is physically far more significant than the position operator in that it governs the time dependent evolution of the system. The Schrodinger equation $\mathrm{i}d\psi/dt=\boldsymbol{H}\psi$ is only available for those $\psi(x)$ in the domain of $\boldsymbol{H}$. If all systems are governed by this evolution then we should clearly hold that only such $\psi(x)$ as are in Dom($\boldsymbol{H}$) represent states. This will clearly exclude functions that are admitted as states by postulate 1. We have good reason, therefore, to believe that $L^2$ spaces are strictly too large-they contain normed functions which cannot be realized as states of any system, contrary to postulate 1.[14] [15, p.533]

This problem is easy to resolve. When $\boldsymbol{H}$ is an unbounded self-adjoint operator, though it is not defined on the whole Hilbert space, its spectral decomposition's projection-valued measure spans the entire space. Accordingly, $\boldsymbol{H}$'s spectral decomposition yields a one-parameter group of unitary operators exp(-i$\boldsymbol{H}t$), which are bounded operators defined on the full Hilbert space. The Schrödinger equation should be replaced with the unitary evolution

$$\psi(t)= \exp(-\mathrm{i}\boldsymbol{H}t)\psi(0).$$

The Hamiltonian operator $\boldsymbol{H}$ is the infinitesimal generator of exp(-i$\boldsymbol{H}t$), and is not necessarily bounded. When $\psi(t)$ is in the domain of $H$, the Schrödinger equation can be obtained as a special case of the unitary evolution according to Stone's theorem [14].

Adopting the unitary evolution exp(-i$\boldsymbol{H}t$) has an advantage over the

[14] Heathcote's postulate 1 corresponds to the axiom (I) in this paper.

Schrödinger equation: it is fully deterministic and is well defined for any time and initial state. Earman [9] takes this as the positive feature of quantum mechanics that avoids the singularities in classic Newtonian mechanics[15]. In contrast, the Schrödinger equation of a system may cease to apply under certain initial conditions after a finite period of time. Thus, it is unnecessary, even undesirable, to restrict quantum states to the domain of the Hamiltonian.

Heathcote provides a further argument, claiming that an infinite amount of energy is required for preparing states with infinite expectation values of the Hamiltonian:

> Now our function is not in the domain of $\boldsymbol{H}$ […] and this means that the expectation for the energy is undefined. But if the product $(\psi, \boldsymbol{H}\psi)$ is not just undefined but infinite for some $\psi$ in $\mathcal{H}$, as it will sometimes be, then we have states which would require an infinite amount of energy to prepare. Since an infinite amount of energy would be required to prepare such states the functions for which $(\psi, \boldsymbol{H}\psi)$ is infinite are physically impossible. [15, p.533]

The generalised expectation value $\langle \boldsymbol{H} \rangle_\psi$ can be defined on certain states outside Dom($\boldsymbol{H}$) in a similar way as for the position operator discussed in Section 2. Having an infinite generalised expectation value of $\boldsymbol{H}$ is a strictly stronger condition than falling outside the domain Dom($\boldsymbol{H}$). Therefore, even if Heathcote is correct, it is insufficient to establish that all states outside Dom($\boldsymbol{H}$) are unphysical.

Heathcote's argument appealing to the preparation of quantum states implies a pragmatic reading of physicality, which I will discuss together with Carcassi *et al.*'s view in Section 4. From the ontological reading of physicality, not all states in reality are the results of the preparation in laboratories by agents, and it is possible that the universe may admit quantum states with infinite expectation values of energy. It is conceivable that a physicist may impose the finiteness of the expectation value of energy as a physical condition grounded in reasonable motivations, like the various energy conditions in relativity. If this is the case, it can find a place in the multiple layers in the hierarchical structure of physicality, which I will discuss in Section 7. Nevertheless, to avoid confusion, I will not introduce this rich structure of physicality at this stage.

[15] As Earman has noticed, Newtonian mechanics is not fully deterministic. I have repeated some of Earman's examples in Section 2.

In the beginning of this section, I have used the plural form "Hamiltonian operators". When Heathcote discusses the Hamiltonian of a particular system, the singular form is used. It should be emphasised that the Hamiltonians for different systems are usually different. If we want to take Heathcote's argument regarding the Hamiltonians seriously, there should not be a single set of all physical states in quantum mechanics. Instead, whether a quantum state is physical depends on the Hamiltonian of the system. This would not pose a serious challenge to the task of determining the set of physical quantum states. A state can consequently be defined as a pair ($\psi$, $\boldsymbol{H}$) rather than $\psi$ alone. In Section 5, we will see that Carcassi *et al*. have not considered this complexity.

## 4. Defining physicality: ontological and modal, or pragmatic and operational?

So far, I have adopted an ontological and modal reading of physicality. From this view, a state is physical if and only if it is possible based on the standards and criteria of physics. By "standards and criteria of physics", I do not mean any assumptions, axioms, or lemmas of any specific physical theories. What I have in mind are some more objective rules that may be called the "laws of nature". The laws of nature may include some more fundamental principles that are not included in or cannot be derived from the axioms of the specific physical theories we work with, such as principles regarding causation or locality. Therefore, there can exist unphysical states that fall within the state space of some specific physical theory (such as the Hilbert space in quantum mechanics).

Although Heathcote does not provide his own definition of physicality, his definition of the "incompleteness" of physical theories is directly relevant:

> Whatever the meaning assigned to the term complete, the following

> requirement for a complete theory seems to be a necessary one: every element of the physical reality must have a counterpart in the physical theory. […] What one requires is that the physical domain that the theory is intended to cover is represented by counterparts in the theory. […] A theory may also be said to be explanatorily incomplete when it goes beyond physical reality but in which there is no procedure for generating the necessary restrictions. (Perhaps an example of this would be the Special Theory of Relativity's allowing for the existence of tachyons, however in this case one would be inclined merely to add to the theory the postulate that super-luminal velocities are everywhere uninstantiated.) [15, p.525]

Heathcote's definition may be summarised as follows: a physical theory is incomplete if either it fails to have a counterpart of (a possible state of) physical reality, or it has a state that has no counterparts of (possible states of) physical reality. It would be appropriate to call those states that fail to have counterparts of (possible states of) physical reality "unphysical".

It should be noted that Heathcote does not use the notion "possible states" in the discussion of incompleteness here. Instead, Heathcote uses the expressions "element of the physical reality" and "everywhere uninstantiated", which implies a form of actualism, according to which a physical theory only needs to have counterparts of what we have in the single *actual world*. The notion of possibility does not appear in the passage I cite. I think Heathcote is not careful enough to distinguish the actualism from a more comprehensive view that a complete physical theory must have and only have counterparts of the physically possible states. In fact, the latter assumption is used in Postulate 1 in Heathcote's paper, which corresponds to the axiom (I) in this paper. Therefore, Heathcote effectively relies on the definition I summarize in his arguments. I will discuss the actualist view that physics should only discuss the single actual world in Section 7. In fact, the actual world itself is also a physically possible world.

Carcassi *et al.*'s formal definition of physicality goes as follows:

> A mathematical definition is physical if it properly characterizes the physical system in the given conditions. That is, if we are able to

> justify on physical grounds that the given mathematical definition is needed to capture and only capture a particular aspect of the physical system under the given conditions. On the other hand, a mathematical definition is unphysical if it can be shown to *require properties or operations that cannot have a physical counterpart*. […] Ultimately, the justification cannot be solely in terms of mathematical matters, but rather it has to rely on that elusive "physical intuition." In fact, the task is exactly to make that elusive "physical intuition" precise enough to turn it into mathematical definitions. This is the job of a physicist. In our specific case, the core argument we will use is that states described by measurable quantities with infinite or pathologically-defined expectation values are *not physically realizable*.[16] [3, p.2]

In other words, unphysical states are those states that cannot have (possible) physical counterparts or are not physically realisable. So far, neither of their definitions has been inconsistent with mine. The word "realisable" has the same etymology with the word "reality". Both terms originate from the Latin word "*realis*". Something is realisable if and only if it can exist in reality. Up to this point, neither of their views has contradicted the ontological and modal reading of physicality in my paper.

However, neither of them fully and consistently concurs with this ontological and modal reading of physicality. As I have shown in Section 3, Heathcote implicitly assumes that any state that cannot be prepared is not physically possible. This does not naturally follow from the ontological reading of physicality since not all states in reality are the result of preparation procedures; rather, it reflects a pragmatic and operational reading of physicality.

After giving the above definition of physicality, Carcassi *et al*. immediately continue that "it doesn't make sense experimentally" and "it can't be done, not even in principle" would be the best argument to show that certain states are unphysical. They also provide the following informal characterisations of "quantum states" and "states in physics", wherein their pragmatic and operational reading of physicality is more

[16] The highlights are added by myself.

explicitly expounded:

> We are only interested in the question of mathematical representation of quantum states, which are physical entities we prepare in a lab. States in physics are the output of a preparation procedure, a characterization that works for both "pure states" of mechanics as well as ensembles in statistical mechanics. They are what we produce, manipulate and, in the end, measure. [3, p.4]

This can be taken as the informal definitions of quantum states and states in physics in general.[17] Moreover, the term "realisable" can also have a more restrictive meaning. Something is realisable if and only if it can be created, produced, crafted, or achieved by human agents under practical circumstances. Compared with the passage cited above, this is probably what Carcassi *et al*. have in mind.

We can flesh out a more complete pragmatic and operational definition of physicality that is not explicitly given in Carcassi *et al.*'s paper:

> A state is physical *iff* it can be prepared, produced, manipulated, and measured by human agents in laboratories.

This definition can be weakened by generalising the notion of "human agents" or "laboratories". However, it is nevertheless distinct from, and strictly stronger than the ontological and modal definition provided earlier in this paper. [18] There can be states that are ontologically possible, but cannot be manipulated or prepared. Moreover, Carcassi *et al.*'s writing implies that "states in physics" should only include the physical states defined above. Although this definition would be very helpful in experimental physics, I will argue that it is inappropriate for addressing the problems that motivate one to find a more precise boundary between physical and unphysical states.

[17] There are actually many readings of this passage. If we take this passage literally, the first sentence, if not merely demonstrating the authors' interests, is demarcating the scope of their paper. It is also unclear whether the second sentence, together with the third, is a definition or an assertion on what states in physics are. Moreover, the notion "states in physics" used in the second half of this passage is more encompassing than mere "quantum states". However, the literal readings do not fit the whole structure of their paper, and it is more coherent to adopt this reading.

[18] If, by "preparation" and "production", we refer to any procedures that can be done in laboratories, even only in principle, then the condition that a physical state can be prepared and produced is not restrictive at all. Quantum states are not created *ex nihil*. The system to be prepared already has a state prior to preparation or production, and it may be in any state that is ontologically and modally physical. Moreover, the null

First, if the above definition applies to all "states in physics", rather than merely quantum states, it would exclude many meaningful states in the research of general relativity and (classical and quantum) cosmology.[19] The research of global spacetime structure, black holes, singularities would be deemed unphysical, as the states (the spacetimes) employed in these studies cannot be prepared and manipulated, even in the most charitable sense. The operational definition of physicality does not fit the research of general relativity. If the definition of physicality has excluded a significant part of physics research, the appropriateness of the definition should be questioned. The inappropriateness of the definition cannot be alleviated by conceding that it only applies to quantum states.[20] It is dubious why quantum theories are so special that they require a different standard of physicality than spacetime theories.

Second, let us revisit the motivations for distinguishing unphysical states from physical states. Carcassi *et al.* argue that the problem of justifying the use of certain mathematical structures in physics is not a mathematical problem, but should be decided on physical grounds. They

---

preparation, which is simply doing nothing, is also a kind of preparation, just as, in a similar fashion, the empty set is also a set. In this reading, no ontologically and modally physical states cannot be prepared. This is likely not the notion of preparation that Carcassi *et al*. have in mind. One may suggest that a preparation in physics must be repeatable and reliable so that it can yield the same states stably. In this reading, the above example does not meet the criteria of preparation, as it depends on the very specific properties of the initial state of the system.

[19] In a response to the earlier manuscript of this paper, Carcassi comments that my example that *a spacetime is physical iff it satisfies Einstein's field equations* (when one does not add on further restrictive principles regarding global causal structures and energy conditions) does not satisfy their definition of physicality. I believe that it is reasonable to speculate that Carcassi would not consider many states in general relativity as physical.

[20] As the first half of the above passage by Carcassi *et al*. is only about "quantum states".

list a series of related questions that should not be decided solely and directly from the mathematical descriptions, but should be judged based on physical considerations. For example,

> Should we really treat continuous and discrete spectra the same, or should we consider that they are, on physical grounds, two rather different things? [3, p.3]

They also point out the cost of failing to provide a clear boundary between physical and unphysical states:

> The cost of not requiring a clear and tight connection between the math and the physics is all around us: an ever-increasing number of interpretations of quantum mechanics, abstract mathematical work whose connection to physics is unclear, theoretical physics more and more disconnected from experimental physics. (*ibid*.)

In this passage, the first cost is the existence of "an ever-increasing number of interpretations of quantum mechanics". I would like to further elaborate on this problem. The mathematical structure of a theory may contain states that possess problematic properties and are difficult to interpret. This henceforth gives rise to pseudo interpretational problems and obscure the real problems of the physical meaning of the mathematical structure. If we can exclude these unphysical states in the mathematical formalism, these pseudo problems and difficulties can be avoided, and we would not need that many interpretations.

However, if we identify certain states as pragmatically unphysical in the mathematical formalism of a theory, interpretational problems would still persist. Most interpretational problems are not about manipulations or observations, but about the underlying ontological structures or what Bell calls "beables", especially in quantum theories. Taking Bell's inequality as an example, which raises probably the most famous interpretational problem in quantum mechanics, it is about the difficulties of determining underlying hidden variables that satisfy certain restrictions. Multiple interpretations are proposed, for example, by endorsing non-locality or non-contextuality. Examining whether these states are pragmatically physical (I believe that the states in Bohmian mechanics would not be) is irrelevant to addressing the interpretational problem.

In addition, the problem regarding whether "on physical grounds" the continuous and discrete spectra are "two rather different things" is an ontological problem regarding the sameness of entities or properties. A pragmatic and operational criterion of physicality is at best insufficient for addressing such a problem.

I do not exclude the possibility that, from a thoroughly pragmatic and operational view of physics, the sameness and differences of things can be translated as completely operational. I also do not deny that the problem of "theoretical physics more and more disconnected from experimental physics" can be addressed by carefully distinguishing which states in theoretical physics can be manipulated and measured. To make the preference of the definitions of physicality more sound in this paper, I will conclude this section by analysing another argument by Carcassi *et al.* on the physicality of states outside the domain of the position operator.

I will return to the argument mentioned, but not fully discussed, at the end of Section 2. This argument is not explicitly stated in Carcassi *et al.*'s paper, but only implicitly conveyed in a passage in the appendix. They revisit the group of unitary transformations discussed in Section 2 which would turn states with finite expectation values for certain observables into states with infinite expectation values, and summarise its consequences:

> In particular, whether the statistical properties of an observable are well defined, or whether it is possible to identify a state through tomography experimentally, would be coordinate/representation dependent. [*ibid.*, p.17]

Needless to say, this would be undesirable and problematic for Carcassi *et al.*. This implies that states with infinite expectation values of certain observables (Carcassi *et al.* have not stated which observables, but the only examples they discuss are the position operator $\boldsymbol{x}$ and its moments $\boldsymbol{x}^n$) cannot be identified through tomography experimentally. Therefore, according to their definition of pragmatical physicality, these states cannot be measured and are thus unphysical.

I will argue that the demarcation between those that can and cannot be identified through tomography cannot be drawn. First, determining a quantum state requires infinitely many degrees of freedom [21] and quantum tomography can at best identify the state approximately. Even if infinitely many degrees of freedom can be obtained, we can at best strictly identify the wave functions of quantum states in a finite (bounded) region. However, the asymptotic behaviour of the quantum states at spatial infinity is still left unspecified.[22] Therefore, the demarcation cannot be drawn.

One may concede that, by "identification of quantum states", one only refers to an approximate identification sufficient for practical purposes. For example, if a wave function (quantum state) is sufficiently localised within a bounded region and its values outside the region can be neglected, it can be said to be "identifiable through tomography".

However, this criterion still fails to classify those states with infinite expectation values of, for example, the position operator as unphysical. In fact, states outside $\mathrm{Dom}(\boldsymbol{x})$ can be nonetheless sufficiently localised within a bounded region. Consider any wave function $\psi(x)$, due to the normalisation of the wave function, the integral of the squared modulus $\psi(x)\overline{\psi}(x)$ outside a bounded region is always finite. When the bounded region is large enough, the "tail" outside the region can be neglected. Moreover, most interestingly, the wave function (quantum state) can be sharply concentrated in a rather small bounded region, but still fall outside the domain $\mathrm{Dom}(\boldsymbol{x})$, as it is the asymptotic behaviour of the quantum states at spatial infinity that determines whether they fall within the domains of position operators, or whether they have infinite expectation values. The modulus amplitude of the wave function outside this sharply bounded region can be extremely small, so that it should be considered "sufficiently localised" to be identifiable through tomography, if any state can be considered identifiable.

[21] As I have mentioned in Section 1, I only consider the infinite-dimensional Hilbert space in this paper.

[22] It is in principle possible that the exact quantum state can be identified if we can identify the wave function within the bounded region over a finite period of time. However, due to the impact of measurement on the system, this cannot be done even in principle.

In addition, as I have already shown in Section 2, falling outside the domain of the position operator and having an infinite (generalised) expectation value of position are two separate conditions. Carcassi *et al*. are rather ambiguous as to which condition decides the identifiability. In conclusion, Carcassi *et al*.'s claim that states outside Dom($\boldsymbol{x}$) are unphysical should be rejected, regardless of whether one adopts the ontological and modal reading or the pragmatic and operational reading of physicality. Furthermore, it remains obscure how the pragmatic definition of physicality can be coherently applied to quantum states.

I do not exclude the possibility that certain arguments can be found for the unidentifiability of certain quantum states. Besides, one may argue that the physicality of states, from the pragmatic and operational perspective, comes in degrees. Some states are clearly physical and some are clearly unphysical, while there are intermediate states.[23] More work will need to be done to support these claims. In the rest of this paper, I will only adopt the ontological and modal reading of physicality and work on developing a more sophisticated account.

## 5. The Schwartz spaces: the sets of physical states?

Heathcote argues that the formalism of quantum mechanics is incomplete when we do not have "the appropriate restrictions" of the Hilbert space [15, p.533]. Therefore, to complete the formalism of quantum mechanics, one has to give another set of physical states smaller than the Hilbert space. Heathcote does not complete this task.[24]

Such a task would not be an easy one. Both Heathcote and Carcassi

---

[23] The *minimal* objective of Carcassi *et al*.'s paper is to demonstrate that at least there are certain states in the Hilbert space that are clearly unphysical. However, in this paper I have argued that even this minimal claim should be rejected.

[24] In fact, it is not clear how seriously Heathcote takes this task. It seems that Heathcote is content with the final negative conclusion that "we should take it as further evidence that we are not yet in possession of the correct mathematical formalism for micro-systems" [15, p.534].

*et al*. believe that physical states must be within the domains of certain operators. However, Carcassi *et al*. have already noticed that no quantum states can fall within the domains of all self-adjoint operators of the infinite-dimensional Hilbert space [3, p.18]. In other words, for any quantum state, it must fall outside the domain of some self-adjoint operator. Moreover, if we require that the quantum state is within the domains of all self-adjoint operators that are the functions of the position operator in the form $f(\boldsymbol{x})$, the resulting space consists of wave functions with compact support on $\mathbb{R}$ only[25]. This space is not closed under even the simplest free evolution. In fact, if $\psi$ has compact support, the quantum state $\exp(-i\boldsymbol{H}_0 t)\psi$ is no longer compactly supported for any $t\neq0$, where $\boldsymbol{H}_0$ denotes the free Hamiltonian operator [19, p.64].

Therefore, one must decide which operators are more important such that any physical quantum states *must* fall within their domains, while the other operators are not so privileged. The question of determining the set of physical states has been transformed into the question of determining the set of *physical observables*. The set of physical states must be within the domains of all physical operators. In fact, Heathcote has considered the option that the set of physical states is the intersection of certain physical operators [15, p.533]. The more observables one considers physical, the fewer states are physical, and *vice versa*. Moreover, the set of physical states must be closed under physical dynamics. If a physical state can evolve to an "unphysical state", such a state should no longer be considered physical. In conclusion, the three sets, namely the set of physical states, the set of physical observables, and the set of physical dynamics, restrict each other, and make the determination quite complicated and ambiguous.

Carcassi *et al*. go further than Heathcote. They suggest that the Schwartz spaces can be a "reasonable alternative" to the set of physical states, which have many desirable features [3, p.7]. Correspondingly, the set of physical observables is spanned by the polynomials of the position and the momentum in the form $\boldsymbol{x}^n\boldsymbol{p}^m$. The Schwartz space $\mathcal{S}(\mathbb{R})$ is a proper subspace of the Hilbert space $L^2(\mathbb{R})$[26]. On the other hand, other operators

---

[25] Again, I only discuss the one-dimensional case for simplicity.

[26] This is the case for the one-dimensional spatial space. The Schwartz spaces for different spatial dimensions are different from each other. Therefore, I have used the

like $e^x$ and $e^p$ may have infinite expectation values on states in $\mathcal{S}(\mathbb{R})$. Carcassi *et al*. do not conclude that this is the ultimate solution. Nevertheless, they still believe that "the properties of Schwartz spaces may provide guidance for a possible comprehensive solution" [*ibid*., p.8]. I will argue that the Schwartz spaces cannot satisfactorily meet the mutual restrictions among physical states, observables, and dynamics without ambiguity and arbitrariness.

First, I will argue that the choice of the set of physical observables is dubious. According to their proposal, $<\boldsymbol{x}>_\psi$ must be finite while $<e^x>_\psi$ can be infinite. Yet they are measured by the same procedure, as any measurement of $\boldsymbol{x}$ is by itself a measurement of $e^x$, and *vice versa*. In a simplified experiment setting, suppose that the apparatus yields a raw datum $\xi_i$ for each measurement. Then, the values of $\boldsymbol{x}$ and $e^x$ can be expressed as functions $A(\xi_i)$ and $B(\xi_i)$ respectively. The expectation values $<\boldsymbol{x}>$ and $<e^x>$ are obtained approximately as $(A(\xi_1)+\ldots+A(\xi_n))/n$ and $(B(\xi_1)+\ldots+B(\xi_n))/n$. Unless the expectation $<\boldsymbol{x}>$ is employed in some fundamental physical process (as in the Hadamard condition discussed in Section 2), allowing divergence for the latter while forbidding it for the former is arbitrary.

Moreover, the convergence of $<\boldsymbol{x}>_\psi$ and other expectation values relies on $\psi$'s behaviour at spatial infinity, and it is unlikely that such behaviour is physically significant. I have already discussed this concern earlier in Section 2. Therefore, discriminating between the operators $\boldsymbol{x}$ and $e^x$ is dubious. One may argue that the restriction is to ensure that physical quantum states are sufficiently localised within a bounded region and to avoid "tails" at spatial infinity.[27] However, I have already argued that quantum states outside the domain of the position operator may still be considered "sufficiently localised". Needless to say, the restriction employed here is stricter and would exclude more states that are sufficiently localised.

---

plural form "Schwartz spaces" earlier in this section. Carcassi *et al*. take this as a desirable feature over the Hilbert space.

[27] In a response to a previous manuscript of my paper, Carcassi explains that the restriction is to ensure that "the tails are 'not fat'".

Second, the Schwartz spaces are not closed under a class of meaningful Hamiltonian evolutions. Therefore, these Hamiltonians must be classified as unphysical.

For simplicity, I will only consider one-particle systems. In quantum mechanics, the Hamiltonian operator of a system is

$$\boldsymbol{H}=\boldsymbol{p}^2+\boldsymbol{V}(x)$$

where $\boldsymbol{p}$ and $\boldsymbol{V}(x)$ are the momentum operator and the potential operator respectively. [28] Strictly speaking, $\boldsymbol{H}$ is usually unbounded and the aforementioned definition has to take into consideration the intricate relations between the domains of operators. In a more rigorous approach, one has to first define an operator $\boldsymbol{p}^2+\boldsymbol{V}(x)$ on the space of infinitely differentiable functions with compact support $C_c^\infty(\mathbb{R}^3)$ instead. This operator is usually not the self-adjoint Hamiltonian operator needed in this section. If this operator is essentially self-adjoint, its closure $\boldsymbol{H}$ is a (usually unbounded) self-adjoint operator on the Hilbert space $L^2(\mathbb{R}^3)$ [6, 9].

Let us now consider what restrictions $\boldsymbol{V}(\boldsymbol{x})$ should have to ensure that $\mathcal{S}(\mathbb{R}^3)$ is closed under the unitary evolution $\exp(-i\boldsymbol{H}t)$. According to the Ehrenfest theorem, the time derivate of the expectation value of an operator $\boldsymbol{A}$ satisfies

$$\mathrm{d}\langle\boldsymbol{A}\rangle_\psi/\mathrm{d}t=\mathrm{i}\langle[\boldsymbol{H},\boldsymbol{A}]\rangle_\psi.$$

If $\boldsymbol{V}(x)$ is a polynomial of (the components of) $\boldsymbol{x}$, then for any polynomial $\boldsymbol{A}$ of position and momentum, $[\boldsymbol{H},\boldsymbol{A}]$ is also such a polynomial. Consequently, $\mathrm{d}\langle\boldsymbol{A}\rangle_\psi/\mathrm{d}t$ is finite, and it follows that $\exp(-i\boldsymbol{H}t)\psi\in\mathcal{S}(\mathbb{R}^3)$.

However, the closedness of $\mathcal{S}(\mathbb{R}^3)$ breaks down when $\boldsymbol{V}(\boldsymbol{x})$ is not a polynomial of (the components of) $\boldsymbol{x}$. For some negative power potentials, $\boldsymbol{p}^2+\boldsymbol{V}(x)$ is still essentially self-adjoint [6]. Some of these potentials are physically meaningful, including the Coulomb potential. We have

$$\mathrm{d}\langle\boldsymbol{p}\rangle_\psi/\mathrm{d}t=\langle-\partial\boldsymbol{V}/\partial x\rangle_\psi.$$

---

[28] In this paper, I have been using the position and the momentum operator on the one-dimensional space only for simplicity. However, one would argue that we do not live in a one-dimensional space and excluding certain Hamiltonians on the one-dimensional spatial space would not be problematic. Therefore, these passages only deal with quantum systems of three-dimensional spatial space $\mathbb{R}^3$. The operators $\boldsymbol{x}$ and $\boldsymbol{p}$ are now considered as three-dimensional vectors. The operator $\boldsymbol{p}^2$ is the inner product of $\boldsymbol{p}$. In the next section, I will use the notations on the one-dimensional space again.

Since $\partial \boldsymbol{V}/\partial$x is not a polynomial of (the components of) $\boldsymbol{x}$, d$<\boldsymbol{p}>_{\psi}/$d$t$ may diverge for some $\psi \in \mathcal{S}(\mathbb{R}^3)$, and exp(-i$\boldsymbol{H}t$)$\psi$ may depart from the Schwartz space $\mathcal{S}(\mathbb{R}^3)$.

In a response to my previous manuscript of this paper, Carcassi explains that the Coulomb potential and the point-like charge are only idealisations, which are "never realised in a physical system". I may infer that, in more realistic systems, the Coulomb potential would be "smeared" and the resulting potential would still be, or at least approximated by a polynomial of (the components of) $\boldsymbol{x}$.

I am unconvinced by this argument. As far as we know, electrons have no internal structures in modern physics. The point-like charge is, in fact, more realistic than one imagines. On the other hand, the continuous distribution of electric charges that avoid the singularities of the potential is instead an idealisation. It is logically consistent for one to claim that the Coulomb potential is unphysical and that the Schwartz space forms the set of physical states. However, it would be too rash to do so at the level of quantum mechanics.

At the end of Section 3, I have suggested that whether a quantum state is physical may depend on the Hamiltonian of the system, and a state should be defined as a pair ($\psi$, $\boldsymbol{H}$) instead of $\psi$ only. According to this definition, if the potential of the system is a polynomial of (the components of) $\boldsymbol{x}$, the relative set of physical $\psi$'s may be given by the Schwartz space.[29] If the potential of the system is the Coulomb potential, the relative set of physical $\psi$'s would be something different. Needless to say, it would be more complicated to give the set of all physical pairs ($\psi$, $\boldsymbol{H}$) under proper restrictions.

In conclusion, the proposal that the Schwartz spaces form the spaces of physical states has serious limitations. It cannot satisfactorily demarcate between physical and unphysical observables, and between physical and unphysical Hamiltonians alike. Carcassi *et al*. have listed many good mathematical properties of the Schwartz spaces (for example, it is closed under the Fourier transform) [3, p.8]. However, as they themselves have argued, the justification of what is physical "cannot be solely in terms of mathematical matters" [*ibid*., p.2]. These features are insufficient to show that they consist of physical, and only physical, states. I am pessimistic regarding whether the ambiguity and arbitrariness of determining the set of physical states, as reflected in the case of Schwartz spaces, can be overcome.

[29] To avoid confusions, I do not call $\psi$'s the "physical states" here, as this term is used to denote the pair ($\psi$, $\boldsymbol{H}$) in this paragraph.

Finally, one may suggest that the ambiguity between physical states and unphysical states cannot be avoided. Some states would be physical without doubt while some states are between physical and unphysical.

I concur with this view. In Section 7, I will argue that physicality should be considered a layered notion that admits degrees. I will develop this view in the final part of the paper.

## 6. Further problems regarding the self-adjointness

This section is relatively independent and readers seeking the main ontological argument may skip directly to Section 7. In this section, I will address the concern raised in the literature regarding whether operators on smaller mathematical spaces can still be self-adjoint.

In critiquing Carcassi *et al.*'s proposal, Lemos [17] worries that self-adjointness of operators cannot be defined on $\mathcal{S}(\mathbb{R})$. Therefore, it is questionable whether the spectral theorem, which is crucial for quantum dynamics, can be reconstructed. Lemos questions whether physical observables can be intrinsically defined as essentially self-adjoint operators on $\mathcal{S}(\mathbb{R})$ without invoking Hilbert space structure:

> It is hard to envisage how the notion of essential self-adjointness can be defined without referring to some extension of the operator's domain, which would inevitably involve going beyond Schwartz space. [*ibid*., p.8]

Even if this can be done, Lemos continues,

> One of the basic tenets of quantum mechanics could be rephrased to state that to each measurable quantity there corresponds an

> essentially self-adjoint operator. It remains to be seen whether this is a fruitful line of inquiry. [*ibid*.]

Heathcote [15, p.529] similarly expresses the concern that, as many physical operators are no longer self-adjoint when restricted to narrower domains, the fundamental postulate of quantum mechanics that *observables are represented by self-adjoint operators* would be challenged.

In their pioneering work on axiomatic quantum field theory, Streater and Wightman [28] have already noticed the great difficulties brought about by restricting the domains of operators:

> Having once resigned ourselves to dealing with unbounded operators, we face a couple of practical problems having to do with their domains. … One is led naturally to assume a common dense domain, $D$, for all the unbounded operators in question … Once one has $D$, another problem arises: to what extent do the values of the unbounded operators on $D$ determine them uniquely wherever else they can be defined? This is particularly acute for observables because an operator which is hermitian[30], when restricted to vectors in $D$, might have several different self-adjoint extensions, and to specify a theory one would have to tell which self-adjoint extension is meant. [ibid., pp.90-91]

In this section, I will explicate their worry.

There does not exist a one-to-one correspondence between essentially self-adjoint operators on $\mathcal{S}(\mathbb{R})$ and self-adjoint operators on $L^2(\mathbb{R})$. A self-adjoint operator's restriction on $\mathcal{S}(\mathbb{R})$ must be symmetric but may fail to be essentially self-adjoint. Conversely, any essentially self-adjoint operator on $\mathcal{S}(\mathbb{R})$ uniquely determines a self-adjoint operator on $L^2(\mathbb{R})$. Therefore, the essentially self-adjoint operators on $\mathcal{S}(\mathbb{R})$ determine a proper subset of self-adjoint operators on $L^2(\mathbb{R})$, and we have reasons to believe that all physically "realistic" Hamiltonians belong to this subset. As shown in the last section, most Hamiltonians are defined as the closures of essentially self-adjoint operators $\boldsymbol{p}^2+\boldsymbol{V}(\boldsymbol{x})$ on $C_c^\infty(\mathbb{R})$. These essentially self-adjoint operators all correspond to essentially self-adjoint operators on $\mathcal{S}(\mathbb{R})$, since $C_c^\infty(\mathbb{R})\subset\mathcal{S}(\mathbb{R})$.

[30] In other words, symmetric. See fn. 3.

Other self-adjoint operators on $L^2(\mathbb{R})$ are conceivable as the Hamiltonians of a system. However, they are not uniquely determined by classical Hamiltonians and would break the classical-quantum correspondence. These Hamiltonians are usually too exotic and are thus classified as unphysical.[31] For example, $\boldsymbol{H}_1=\boldsymbol{p}^2-\boldsymbol{x}^4$ defined on $\mathcal{S}(\mathbb{R})$ is symmetric but not essentially self-adjoint [14]. In general, a symmetric but not essentially self-adjoint operator may admit no self-adjoint extensions or infinite self-adjoint extensions [13]. $\boldsymbol{H}_1$ has infinite self-adjoint extensions and none is "canonical". There exists a state $\varphi\in L^2(\mathbb{R})$ such that $(\varphi,(\boldsymbol{p}^2-\boldsymbol{x}^4)\varphi)\neq((\boldsymbol{p}^2-\boldsymbol{x}^4)\varphi,\varphi)$ [14], and therefore any self-adjoint extension of $\boldsymbol{H}_1$ cannot preserve the form of $\boldsymbol{p}^2-\boldsymbol{x}^4$ on certain states even if $\boldsymbol{p}^2-\boldsymbol{x}^4$ can be defined on these states. $\boldsymbol{H}_1$'s classical Hamiltonian counterpart has the exotic feature that particles can be expulsed to spatial infinity within a finite period of time. It should be expected that realistic physical systems should not have increasing or non-diminishing repulsive force at spatial infinity. In fact, for most realistic physical systems, their quantum Hamiltonians, when restricted on $C_c^\infty(\mathbb{R})$, are essentially self-adjoint [22]. If $\mathcal{S}(\mathbb{R})$ is closed under these Hamiltonian evolutions as discussed in the last section, one can continue to use the Stone theorem and spectral decomposition for quantum mechanics on $\mathcal{S}(\mathbb{R})$.

Therefore, Heathcote's worry may be alleviated. Now, I will turn to Lemos's worry. The definition of essential self-adjointness, naïvely, relies on the structure of the Hilbert space larger than the Schwartz spaces. Most criteria of essential self-adjointness also involve operators defined on space larger than the Schwartz spaces (for example, the adjoint of symmetric operators on the Schwartz spaces). Yet, on the other hand, any Schwartz space uniquely determines a Hilbert space as its completion with respect to its norm. I take it as indisputable that the norm is an "intrinsic" property of a Schwartz space. Then, features appealing to the structure of $L^2(\mathbb{R})$, which is the $L^2$-completion of $\mathcal{S}(\mathbb{R})$, can be understood as "intrinsic" features of $\mathcal{S}(\mathbb{R})$ in an indirect way. I suspect whether it is possible to provide a clear account to classify the "intrinsic" features of a mathematical structure such as $\mathcal{S}(\mathbb{R})$. Can "belonging to ($\pi$-1, $\pi$+1)" be understood as a property relying only on intrinsic features of $\mathbb{Q}$? I doubt whether pursuing such an account is necessary for physics research.

[31] The potentials of these Hamiltonians are usually not bounded from below and constitute genuine examples of unphysical dynamics, in contrast to the dynamics with the presence of the Coulomb potential discussed in Section 5.

Furthermore, I suggest that our attitudes towards Lemos's worry depend on how we interpret Carcassi *et al.*'s proposal. Under a more radical interpretation, they replace the received axiomatic quantum mechanics with a new formulation with a set of axioms defined on the Schwartz spaces. Then, the simplicity of these axioms would be questioned if they too often rely on mathematical structures larger than the Schwartz spaces. However, Carcassi *et al.*'s goal is much less disruptive. They only want to distinguish the unphysical quantum states from the physical quantum states and do not radically change the mathematical formalism of quantum mechanics. Actually, they are not reluctant to use the Hilbert space structure for obtaining "physical results":

> When solving an integral, we sometimes extend the domain from real to complex values, which are unphysical, because complex analysis has nicer mathematical features. The result so found, however, is still a valid result in the real domain. Similarly, we can pose a problem on the Schwartz space, extend to the Hilbert space or the space of distributions for calculation, and then bring the result back to the Schwartz space. [3, p.19]

Under this less radical interpretation, we do not redefine the quantum Hamiltonians of systems as essentially self-adjoint, and can continue to use any self-adjoint operators of the Hilbert spaces. Therefore, Lemos's worry should also be alleviated.

## 7. The ambiguity and hierarchy of physicality

In this paper, I have been following the ontological and modal reading of physicality. Following this read, I will not distinguish the notions of "physicality" and "physical possibility". I have provided a tentative

definition in Section 1:

> A state $\psi$ is physically possible *iff* for any proposition $p$ that holds for $\psi$, $p$ is consistent with the laws of nature.

The definition can be slightly modified to accommodate the terminologies used in this paper and in the modal metaphysics of possible worlds:

> A world represented by a state $\psi$[32] is physically possible *iff* it is consistent with the laws of nature.

The above definition does not imply any degree of ambiguity. A state or a possible world either obeys the laws of nature or violates the laws of nature. Furthermore, a similar diagram of Figure 1 showing the nesting relations between the set of physically possible worlds, the set of metaphysically possible worlds, and the set of logically possible worlds is given in the literature.[33] These diagrams also assume that the boundaries of these sets are determinate and admit no ambiguity.

However, the above definition does not provide the guide for determining which states are consistent with the laws of nature. This leaves room for the discretion of physicists and philosophers from different perspectives and criteria. When we use a physical theory to describe reality, we have the set of all states in the mathematical formalism of the theory, denoted by $\Omega$. For quantum mechanics, the set $\Omega$ is identical to the Hilbert space $\mathcal{H}$. As I have discussed in Section 1, not all states in $\Omega$ may be allowed by nature, especially when some states have pathological features and properties.

Among all options of determining the set of physically possible states, there are two extremal views that admit the most or the fewest states as physically possible:

> (Maximalism) All states in the mathematical formalism of the physical theory that we use to describe reality, as denoted by $\Omega$, are physically possible.

[32] In this section, I do not distinguish between a possible world and a possible state. A state is understood as the representation of a world here.
[33] For example, a similar one can be found in [20].

(Minimalism) There is only one physically possible world, which is the actual world.

The latter view also has some supporters in the literature on the foundations of physics, which is called "strong determinism" [5,6]. Chen proposes that the initial quantum state is a mixed state[34] fixed by some principles in the philosophy of statistical mechanics. These principles, including the "past hypothesis", are identified as laws by Chen, and accordingly a quantum state must be consistent with them to be physical. In Section 4, this view is also called "actualism". The maximalist view is assumed in most literature on quantum mechanics.

In some scenarios, the maximalist view may seem too permissive while the minimalist view seems too restrictive. Taking general relativity as an example, the set $\Omega$ of general relativity consists of all four-dimensional Lorentzian manifolds $\mathcal{M}$ with the metric fields $g$ and energy-momentum tensor fields $T$ that obey the Einstein equations. $\Omega$ would include worlds that have pathological causal structures, including the Gödel spacetime, which admits causal loops and non-global hyperbolic spacetimes that violate criteria of determinism. Various restrictions have been proposed that physical spacetimes must obey, including various energy conditions [8], the global causal structures, and the prohibition of naked singularities. For classical mechanics, determinism can be violated even at small scales when the potentials do not satisfy the Lipschitz condition. One famous example in the foundations of physics literature is the Norton's dome [9,21]. To avoid these singular behaviours, one may impose that the potentials in all physically possible worlds must satisfy the Lipschitz condition.

[34] In this paper, until now, I have only discussed pure states in quantum mechanics. Traditionally, mixed states are understood as not being as real as pure states, but merely representations of ignorance. The introduction of mixed states in Chen's papers is motivated by the foundations of statistical mechanics. I will discuss mixed states again in the next section, in a discussion of the ontology of quantum field theory.

Then, we have many options between the extremal maximalism and minimalism. The plausibility of the restrictions in these intermediate options is debated in the foundations of physics literature. For example, Smeenk and Wüthrich suggest that the view that a physical spacetime must be globally hyperbolic is not well motivated [27]. These restrictions may also be rejected empirically. If it is observed that the actual world that we inhabitant violates some restrictions, these restrictions should be rejected as criteria for physicality, for the actual world must be physically possible. Various energy conditions should particularly undergo such empirical tests.

This paper has been discussing various proposals on the restrictions on physical states in quantum mechanics. Unlike in general relativity and in classical mechanics, these restrictions in quantum mechanics are much more ambiguous. Each of these restrictions requires that the physical states must fall within the domains of certain self-adjoint operators, but it is difficult to decide which states must be privileged without arbitrariness. Moreover, I have argued that we do not have sufficient reasons to disregard the physicality of states merely for falling outside the domains of these self-adjoint operators. Therefore, the maximalist view seems suitable for quantum mechanics.

However, although all quantum states in $\Omega$ should be considered as physical from the maximalist view, it is reasonable that Heathcote and Carcassi *et al*. may worry about the peculiar properties of certain states in $\Omega$. Although there are no sufficient reasons to dismiss the physical possibility of those quantum states outside the domain of the position operator, such as the principles regarding determinism or causation, these quantum states indeed do not seem equally as "regular" or "normal" as other states. If the quantum state of the actual world is within the domains of, at least, the first few powers of the position operator and the momentum operator, it is natural to suppose that "normal" quantum states should also fall within these domains. In other words, although all quantum states are physically possible, they are not equally possible.

We can formally define the intermediate views of physicality with certain restrictions as follows:

($\theta$-Intermediatism) A state in the mathematical formalism of the physical theory that we use to describe reality is physically possible *iff* it satisfies the restriction $\theta$.

Each restriction $\theta$ defines a set $\Omega_\theta$, as a subset of $\Omega$, consisting of all states that are physically possible according to the $\theta$-Intermediatist view of physicality. As I have mentioned earlier in this section, the state of the actual world must be included in any such set $\Omega_\theta$.[35] I will also call each set $\Omega_\theta$ a *hierarchy* or a *layer* of physicality. The more hierarchies of physicality a world falls within, the more it resembles the actual world, and the more physically possible it is. On the other hand, the fewer hierarchies of physicality a world falls within, the more remote it is to the actual world, as shown in Figure 2. It follows that physicality is not a univocal concept, but a layered one.

For quantum mechanics, such a restriction $\theta$ corresponds to a set of self-adjoint operators. The set $\Omega_\theta$ is the intersection of the domains of all these self-adjoint operators. The more common domains of self-adjoint operators a world falls within alongside with the actual world, the more physically possible it is. The hierarchies of physicality form a partially-ordered set under the inclusion relation. It is usually not a totally-ordered set. For example, for the two hierarchies of physicality $\mathrm{Dom}(\boldsymbol{A})\cap\mathrm{Dom}(\boldsymbol{B})$ and $\mathrm{Dom}(\boldsymbol{A})\cap\mathrm{Dom}(\boldsymbol{C})$, neither is necessarily be the subset of the other.

Consequently, the degree of physicality of a world is characterised by its similarity to the actual world. This can also be applied to the physicality of states in general relativity and classical mechanics. The restrictions regarding determinism and causal structures, if not justified as *a priori* philosophical principles, can be motivated by the similarities to the actual world. If we find that the actual world is deterministic insofar as we have observed it, it is natural to suppose that possible worlds that satisfy the mathematical structures of our physical theory must also share this feature to be close enough to the actual world, and to be physically possible.

---

[35] If we treat the state of the actual world as evolving over time, then we should require that the state of the actual world *at any time* is included in any such set $\Omega_\theta$. We may also consider the set of possible histories instead of possible instantaneous states, or adopt the Heisenberg picture, to avoid this issue.

The hierarchical view of physicality proposed in this section is consonant with the suggestion at the end of Section 5 that while some states are clearly physical, some other states are between physical and unphysical. Those "clearly physical" states are states with higher degrees of physical possibility, and correspondingly the states between physical and unphysical are states with lower degrees of physical possibility. This should not be surprising from the perspective of modal metaphysics. In modal metaphysics, although a single set of all possible worlds is chosen, the set possesses rich structures that can characterise the degree of possibility in terms of the closeness to the actual world [18].

Finally, some readers may find the continuous discussions of "possible worlds" unsatisfactory. A possible world encompasses the totality of physical reality (in a hypothetical scenario if it is not actual). In physics, we sometimes do not consider the state of all the things, but only the state of a subsystem. For example, in quantum mechanics, we usually do not consider the quantum state of the whole universe (although we do sometimes), but instead the quantum state of some microscopic subsystem. It should therefore be questioned whether the hierarchical view of physicality proposed in this section is disconnected from the real physics research.

I believe that in principle there are no insurmountable distinctions between states of possible worlds and possible states of subsystems. The quantum state of a subsystem can be supplemented by the quantum state of the rest of the universe and their correlations, and henceforth become the quantum state of a complete possible world. The quantum state of the subsystem can thus be identified with a class of possible worlds. Each member of the class is obtained by supplementing the subsystem's quantum state. Formally, the degree of physicality of such a subsystem's quantum state can be characterised indirectly by the respective degrees of physicality of its corresponding class of possible worlds. It would be very complicated to flesh out such an account, but without giving the technical details I believe that it is doable.

Besides, one can take the theory of possible worlds as a useful formal framework for analysing possibility and modality without appealing to the totality of (possible) reality. One only need to replace the occurrence of "possible worlds" with "possible states of subsystems" to give a refined hierarchical account of physicality, with only a few modifications.

One may still suspect the usefulness of the hierarchical view of physicality. As I have argued throughout the paper, in quantum mechanics, no intermediate layers of physicality are privileged over others (otherwise we would not have the inevitable arbitrariness and ambiguity in determining which operators must have finite expectation values). It would be more convenient if we only adopt the maximalist view for quantum mechanics. The hierarchical view of physicality seems to be merely a characterisation of the closeness between possible worlds and the actual world (though it should not be surprising that modal properties can be characterised by similarity). How useful is it for the research of physics? In the next section, I will show that the hierarchical view of physicality is particularly helpful in the foundational discussions of quantum field theory, where no privileged set Ω may exist.

## 8. The hierarchy of physicality and the ontology of quantum field theory

In quantum field theory, we do not even have a consensus on what the mathematical formalism of the theory is, and what states the formalism admits. For a quantum system, a C*-algebra $\mathcal{W}$, known as the Weyl algebra, can be generated by the commutation or anti-commutation relations of the system.[36] $\mathcal{W}$ is a Banach algebra that encodes the minimal observables of the system. However, it is usually too small for physics, as it does not include many physically significant observables, such as the total number of particles and their corresponding spectral projections. In addition, the spectral theorem is also generally inapplicable to the Weyl algebra. Therefore, it needs to be enlarged to include more observables. From the perspectives of both mathematics and physics, it is more desirable if the algebra of all observables of a quantum system could have more topological structures. As a Banach algebra, $\mathcal{W}$ has the structure of norm topology as a Banach space, but it lacks the structures of strong operator topology and weak operator topology. It would be ideal to "concretise" the Weyl algebra by granting it the structure of operators on some Hilbert space. The standard treatment is to represent the algebra $\mathcal{W}$ as $\pi(\mathcal{W})$ on a Hilbert space $\mathcal{H}_\pi$. $\pi(\mathcal{W})$ consists of operators on $\mathcal{H}_\pi$. By resorting to the Hilbert space structure of $\mathcal{H}_\pi$, we can consequently define the weak closure $\pi(\mathcal{W})''$ under the weak operator topology, which is not only a C*-algebra, but also a von Neumann algebra. It is called the von Neumann algebra affiliated with the representation $\pi$ of $\mathcal{W}$ in the literature [24, p.87].

[36] For quantum mechanics, a typical member of the Weyl algebra has the form $e^{i(\alpha\boldsymbol{x}+\beta\boldsymbol{p})}$, where $\boldsymbol{x}$ and $\boldsymbol{p}$ represent the position and momentum as observables, respectively. One should be wary that the Weyl algebra is defined purely algebraically and none of $e^{i(\alpha\boldsymbol{x}+\beta\boldsymbol{p})}$, $\boldsymbol{x}$, and $\boldsymbol{p}$ should be understood as the operators on some Hilbert space. Only when we build the representations of the Weyl algebra, these observables are mapped to operators on some Hilbert space. In quantum field theory, the position and momentum are replaced by observables generated by the *distributions of the observables quantum field* such as $\boldsymbol{\varphi}(x)$ and $\boldsymbol{\pi}(x)$ [18]. The distributions $\boldsymbol{\varphi}(x)$ and $\boldsymbol{\pi}(x)$ correspond to the so-called quantum field operators and the conjugate momentum density operators in the physics literature. However, they are not real operators. As a Banach algebra, all members of the Weyl algebra are bounded (in the algebraic sense).

According to the well-known Stone-von Neumann theorem, all weakly continuous irreducible representations of the Weyl algebra of ordinary quantum mechanics are equivalent to each other. We can, without loss of generality, choose any such representation $\pi$ and its affiliated von Neumann algebra $\pi(\mathcal{W})''$ is identical to the algebra of all bounded operators $\mathcal{B}(\mathcal{H}_\pi)$ on $\mathcal{H}_\pi$. The physically significant unbounded operators, such as the position operator $\boldsymbol{x}$ and the momentum operator $\boldsymbol{p}$, can be obtained as the generators of strongly continuous one-parameter unitary groups consisting of members of $\mathcal{B}(\mathcal{H}_\pi)$. If we identify the set of physical observables with the set of self-adjoint operators in $\mathcal{B}(\mathcal{H}_\pi)$, together with the self-adjoint generators defined above, we will return to the axiom (II) in the beginning of this paper.

For quantum field theory, we can similarly introduce the unbounded self-adjoint operators as the generators of strongly continuous one-parameter unitary groups consisting of operators in $\pi(\mathcal{W})''$. We may need to repeat the discussions on the physicality of quantum states outside their domains. However, the problem with quantum field theory is more than that. The Stone-von Neumann theorem does not hold for quantum field theory, and we may have inequivalent weakly continuous irreducible representations. There is consequently uncertainty in finding the appropriate set of physical observables and realisations in quantum field theory. This is known as the problem of inequivalent representations, and it poses a serious challenge to interpreting quantum field theory. There is enormous literature on this topic in the foundations of physics[37], and it is impossible to offer a complete treatise within this paper's scope. I will only schematically show how the hierarchical view of physicality developed in Section 7 can be used to solve the interpretational difficulties brought by the problem of inequivalent representations.[38]

As Belot [1] summarises, traditionally, to interpret a physical theory is to divide the states "into two sharply defined classes: the physically possible ones allowed by the laws, and the physically impossible ones". In fact, this task cannot be separated from another task of determining the set of all physical observables, in both quantum mechanics and quantum field theory.

With the Weyl algebra $\mathcal{W}$ at hand, the most convenient option is to identify the set of physical observables with $\mathcal{W}$, and identify any algebraic state $\omega$ of $\mathcal{W}$ as physically possible. An algebraic state of a C*-algebra is a normed positive linear functional of the algebra. When the system is in the state $\omega$, the expectation value of any observable $\boldsymbol{A}\in\mathcal{W}$ is $\omega(\boldsymbol{A})$. This option is called "algebraic imperialism" by Ruetsche [24]. As I have discussed earlier in this section, $\mathcal{W}$ is too small. Correspondingly, the set of physical states is too large, as it fails to take into consideration the restrictions imposed by more physical observables.

On the other hand, one can choose a representation $\pi$ of the Weyl algebra $\mathcal{W}$ and realise it on some Hilbert space $\mathcal{H}_\pi$. The set of physical observables can be chosen to be $\pi(\mathcal{W})''$. The set of physically possible worlds is identified with the set of all normal states of $\pi(\mathcal{W})''$. One can similarly define the algebraic states of $\pi(\mathcal{W})''$. However, some of these states may be unphysical. Those σ-additive states are called normal. In other words, a state $\omega$ is normal *iff* for each countable set $\{\boldsymbol{E}_i\}\subset\pi(\mathrm{W})''$ of pairwise orthogonal projection operators, $\omega(\Sigma\boldsymbol{E}_i)=\Sigma\omega(\boldsymbol{E}_i)$ [24, p.90][39]. For quantum mechanics, a state is normal *iff* it can be represented as a density operator on the Hilbert space. However, in quantum field theory, this result does not always hold. Non-normal states of a von Neumann algebra have undesirable features that may render them unphysical. It may undermine some profound logical principles. For example, if the set $\{\boldsymbol{E}_i\}$ represents a series of events, then $\omega(\boldsymbol{E}_i) > 0$ indicates that the event may happen, and the condition $\omega(\Sigma\boldsymbol{E}_i) > 0$ indicates that at least some of the events represented by $\{\boldsymbol{E}_i\}$ may happen. However, if $\omega$ is non-normal, it is possible that $\omega(\Sigma\boldsymbol{E}_i)=1$ while $\omega(\boldsymbol{E}_i)=0$ for any event $\boldsymbol{E}_i$. On the face of it, these states should be excluded from the formalism of quantum theories.

---

[37] For introductions, see [1,7,24-26].

[38] Here I am following the algebraic approach of quantum field theory. There have been debates that it is not the suitable formalism for foundations of physics discussions as it is yet unable to reproduce all the results of the non-rigorous quantum field theory in the physics literature [12,31]. Although the algebraic quantum field theory has its limitations, I believe that it is an appropriate starting point and will be instructive for further foundational analysis.

The option of determining the sets of physical observables and physical states via the von Neumann algebra affiliated with some specific representation $\pi$ of $\mathcal{W}$ is called "Hilbert space conservatism" by Ruetsche, as many familiar results in quantum mechanics can be restored via the Hilbert space structure. However, due to the existence of inequivalent representations, it is debatable which representation, together with its affiliated Hilbert space, is the canonical one.

A natural option is to choose the Gel'fand-Naimark-Segal representation $\pi_\omega$ of some physically significant state $\omega$, such as the vacuum state. Clearly, the state $\omega$ itself is a normal state of the von Neumann algebra $\pi_\omega(\mathcal{W})''$. We define the folium $\mathcal{F}_\omega$ of $\omega$ as the set of all normal states of the representation $\pi_\omega(\mathcal{W})''$. Equivalently, it can be proven that a state lies in the folium $\mathcal{F}_\omega$ *iff* it can be expressed as a density operator on the Hilbert space $\mathcal{H}_{\pi\omega}$ [24, p.96].[40] However, the vacuum state is not

---

[39] The condition that these operators are pairwise orthogonal projections is to ensure the existence of the limit $\Sigma\boldsymbol{E}_i$. The limit is defined via the strong operator topology and relies on the Hilbert space structure of $\mathcal{H}_\pi$.

[40] In the earlier sections of this paper, I have only considered the pure states in quantum mechanics, which correspond to elements of the Hilbert space. Mixed states are usually taken as merely representing ignorance of the actual pure quantum state. However, in quantum field theory, mixed states are more essential and cannot be

unique in quantum field theory. It depends on, for instance, the choice of the global foliation of spacetime [6a, 18]. Regardless of which vacuum state we choose as the physically preferred one, there are always some physically significant states outside the folium of the state we choose. For example, if we choose the so-called Minkowski vacuum that reflects the global symmetry of spacetime as the preferred one, the Unruh vacuum that plays a crucial role in explaining the Unruh effect would lie outside its folium.

One may then ask whether we can choose a "larger" representation that may include both the Minkowski vacuum and the Unruh vacuum as its normal states. This option is possible and is called "universalism" by Ruetsche. However, it is still unsatisfying. Ruetsche criticises this option for admitting an overabundance of observables into the representation. Each observable is represented for "continuously many times" according to Ruetsche [24, p.145].

It is impossible to survey all proposed options and related debates concerning this highly sophisticated problem within the scope of this paper. It suffices to show that the traditional view of interpretating a physical theory has encountered great difficulties. Ruetsche's own solution is to give up the traditional view and to adopt the so-called "coalescence approach":

> It follows that there can be an *a posteriori*, even a pragmatic, dimension to content specification, and that physical possibility is not monolithic but kaleidoscopic. Instead of one possibility space pristinely associated with a theory from the outset, many different possibility spaces, keyed to and configured by the many settings in which the theory operates, pertain to it. Following Kadison, I call this the *coalescence approach* to interpreting physical theories. [24, p.147]

In other words, as Wallace summarises, the interpretation of a physical theory is situation-dependent [32]. To put it in another way, which set of states is physically possible is not unrestricted, but depends on which

---

dispensed with [23].

question we are concerned with. When using quantum field theory to solve different problems, we should identify different sets of physically possible states. Jacobs gives a clearer definition of this situation-dependent account:

> The interpretation of a theory depends on contingent facts about target system's state and on the context in which the theory is used. [16, p.1078]

Jacobs acknowledges that this position would pose a threat to realism.

The hierarchical view of physicality developed in the last section could provide another solution to the interpretational problem of quantum field theory without undermining realism. According to the hierarchical view of physicality, there is not a single set of physically possible states, but a collection of multiple such sets $\Omega_\theta$. We can identify each set $\Omega_\theta$ with the folium $\mathcal{F}_\theta$ of some state $\theta$ (see Figure 3). When interpreting quantum field theory, the state of the actual world is left unspecified. Therefore, we have "many different possibility spaces" as Ruetsche suggests. These possibility spaces (hierarchies of physicality) are fully objective and do not depend on the context in which the theory is applied and used. The degree of physicality of a world depends on the specification of the actual world.

In modal metaphysics, as discussed in the last section, there is a single set consisting of all possible worlds. However, not all possible worlds are accessible to each other. The definitions of the necessity and possibility of a proposition quantify only over all *accessible* possible worlds relative to the actual world. This reveals the indexicality of possibility. Although this idea is well accepted in the modality literature, it is not reflected in most foundations of physics research. For quantum field theory, among all hierarchies $\Omega_\theta$, the folia of pure states are of particular importance. It can be proven that they are either identical or disjoint. Furthermore, they are also the candidates of Ruetsche's Hilbert space conservatism. States that are inaccessible from the actual world, such as the Unruh vacuum when the actual state of the universe is the Minkowski vacuum, remain physically possible in an unquantified sense, yet they do not share essential structures with the actual world. For example, different sets of physical observables may be defined over different sets of mutually accessible states, namely different folia. This is by no means surprising. In modal metaphysics, different sets of mutually accessible possible worlds likewise have different ontological structures. Therefore, the weakness of choosing only one set of physical states can be overcome.

In this section, I only briefly show how the hierarchical view of physicality can be used to address foundational problems in physics. I believe that this proposal deserves further research.[41]

## 9. Conclusion

In this paper, I begin with the problem of the physicality of quantum states outside the domains of certain self-adjoint operators in quantum mechanics, such as the position operator and the Hamiltonian operators. I have argued that we do lack sufficient reasons to disregard them as unphysical from the ontological and modal reading of physicality. Moreover, I have argued that there are ambiguity and arbitrariness if we want to draw a distinction between physical and unphysical states of the Hilbert space. Then, I have proposed a more sophisticated account of physicality, known as "the hierarchical view", that has a closer connection to modal metaphysics. In the final section, I briefly show how the newly proposed hierarchical view of physicality can help to address interpretational problem of quantum field theory.

To sum up, physicality should not be understood as a monolithic concept, but a layered one that admits degrees. The research of the interpretation of quantum field theory in this paper also suggests that some physical theories naturally induce modal structures.[42] The development of modal metaphysics, especially the analysis of counterfactuals and comparative possibilities, has not been frequently reflected in the research of the foundations of physics. I hope that this paper serves as an invitation for further work at the intersection of these fields.

---

[41] Jacobs's own suggestion "the interpretation of a theory depends on contingent facts about the target system's state, but not on the context in which the theory is used" is close to my proposal in this section, but Jacobs does not explore deeper connections between the interpretation of physical theories and modality metaphysics [27].

[42] Butterfield [2] also suggests that the Hamilton-Jacobi formalism of classical mechanics naturally accommodates David Lewis's theory of modality and counterfactuality.

**Data Availability Statement:** The author does not analyse or generate any datasets.

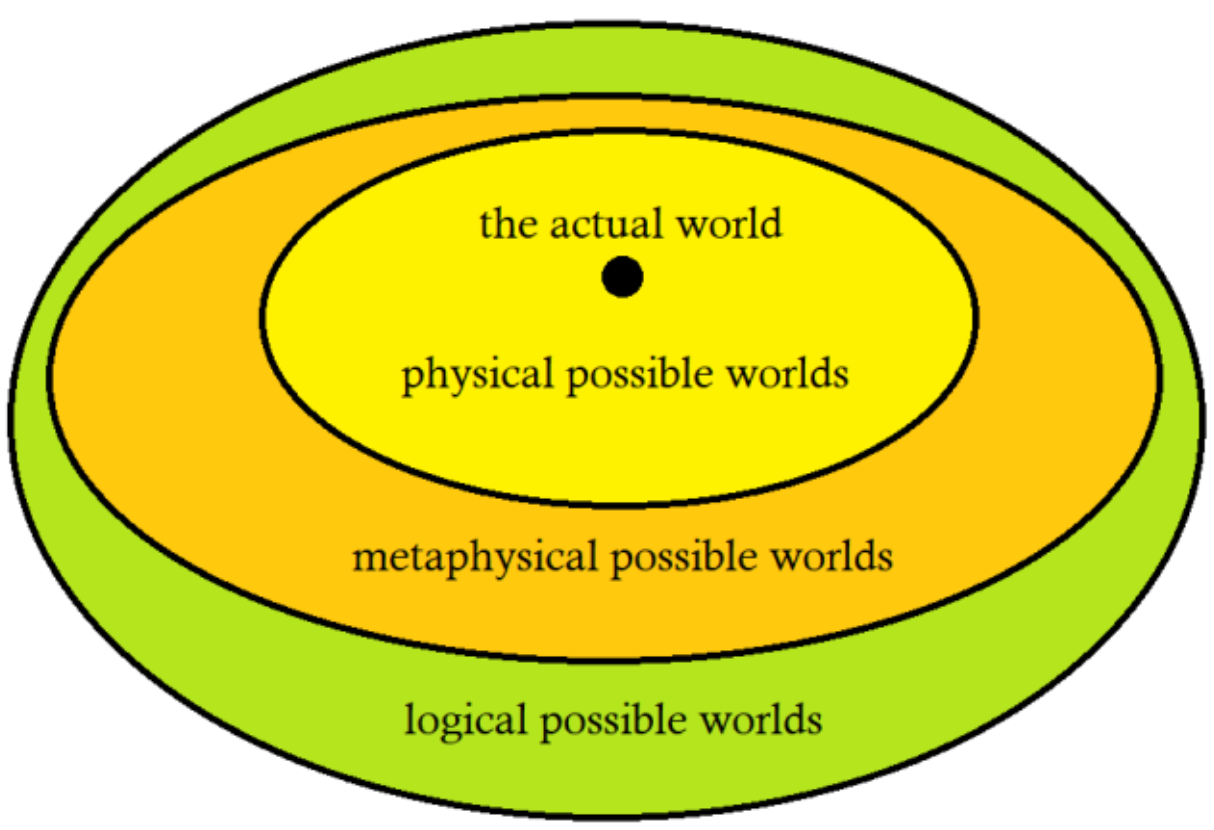

**Figure 1**
The hierarchy of logical, metaphysical, and physical possible worlds

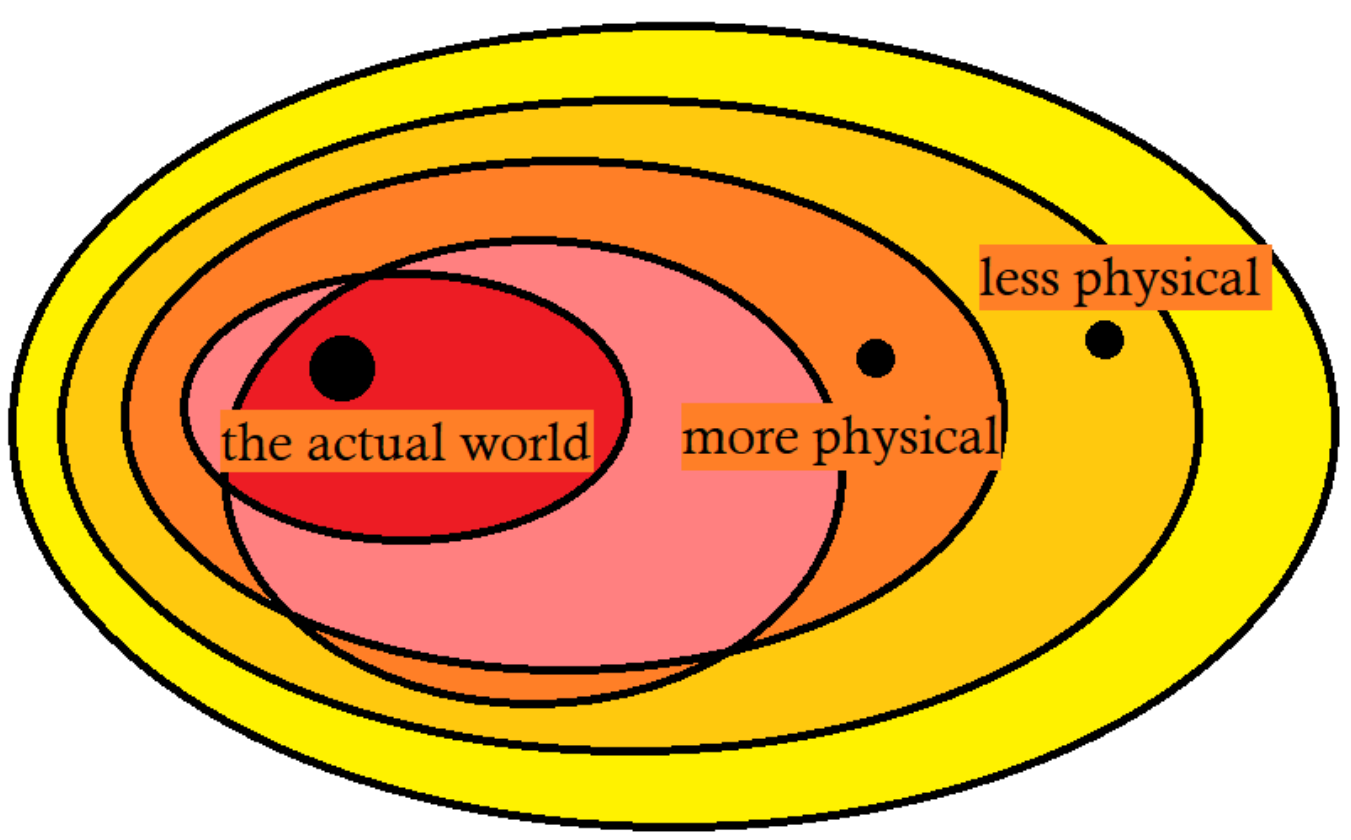

**Figure 2**
The hierarchies of physically possible worlds
(each oval represents a hierarchy)

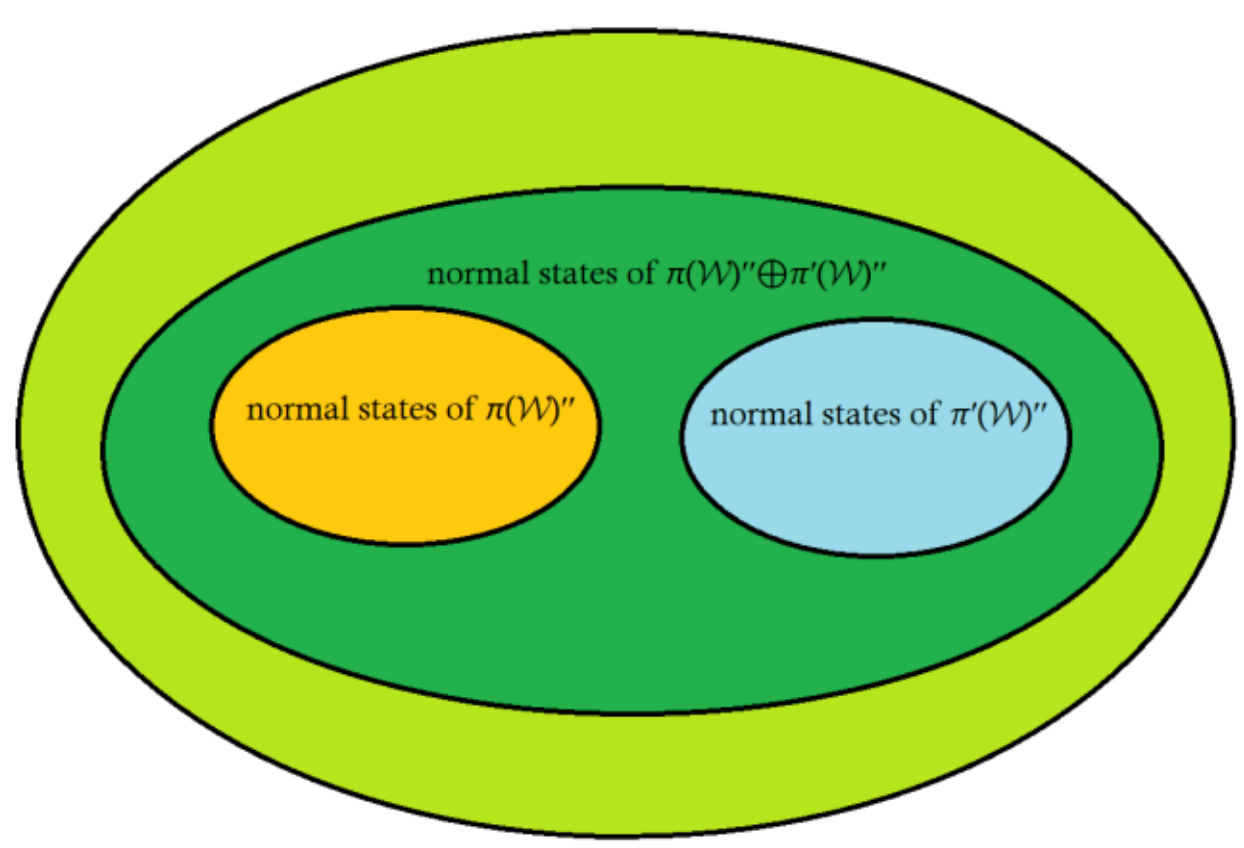

**Figure 3**
The hierarchies of physically possible worlds in quantum field theory